\documentclass[12pt,epsf]{revtex4}
\usepackage[dvips]{graphicx}
\usepackage[dvips]{color}

\begin{document}

\title{The role of cooperative binding on noise expression}
%[Cooperative binding and noise expression]
%\title{The role of binding site cluster on noise expression}
%\title{The architecture of transcriptional switches}

\author{P.S. Gutierrez$^1$,  D. Monteoliva$^2$ and
L. Diambra$^1$\footnote{To whom correspondence should be addressed. E-mail:
ldiambra@creg.org.ar}}
\address{ $^1$Laboratorio de Biolog\'{\i}a de Sistemas -- CREG-UNLP, \\
Av. Calchaqui Km 23.5 CP 1888, Florencio Varela, Argentina. \\
$^2$Instituto de F{\'i}sica -- UNLP \\
CC 67 CP 1900, La Plata, Argentina}

\begin{abstract}
The origin of stochastic fluctuations in gene expression has
received considerable attention recently. Fluctuations in gene
expression are particularly pronounced in cellular systems because
of the small copy number of species undergoing transitions between
discrete chemical states and the small size of biological
compartments. In this paper, we propose a stochastic model for gene
expression regulation including several binding sites, considering
elementary reactions only. The model is used to investigate the role
of cooperativity on the intrinsic fluctuations of gene expression,
by means of master equation formalism. We found that the Hill
coefficient and the level of noise increases as the interaction
energy between activators increases. Additionally, we show that the
model allows to distinguish between two cooperative binding
mechanisms.
\end{abstract}
%\pacs{87.18.Cf, 87.16.Yc, 87.18.Tt}

\maketitle

\section{Introduction}

All chemical reactions have intrinsic fluctuations that are
inversely proportional to the system size. Such fluctuations are
particularly pronounced in gene expression. At the transcriptional
level, gene expression is mainly controlled by the {\it
cis}-regulatory system (CRS) and transcription factor (TF) proteins
that bind specifically to DNA sites \cite{ptashne}. The TFs
influence the transcription rate by interacting with other
transcriptional components (RNA polymerase, TATA-binding protein,
etc.). Like any molecular interaction, the binding of TFs to the
regulatory sites is a stochastic event rendering the transition
between states of the CRS a stochastic process. This source of noise
is known as intrinsic noise in gene expression regulation, to
distinguish it from that produced by other influences such as random
fluctuations in nutrients, cell division or regulatory inputs to the
transcriptional machinery, known as extrinsic noise
\cite{swain,elowitz02}.

There is a broad variety of different CRS motifs that underly such
regulation. The diversity of CRS range from simple ones to more
complex motifs that include dozens of regulatory sites, some of them
organized in clusters or tandems \cite{ptashne}. This cluster
organization points to cooperative effects in the gene regulatory
process, because proteins rarely seem to bind to DNA without
interacting with other DNA-binding proteins. Despite this
complexity, the bulk of stochastic models for gene regulation are
based on transitions between two promoter states (active and
inactive) and, recently, more complex models have been explored
\cite{blake,pnas08}. All these models approximate the
transcriptional control by using a regulatory expression function
(Hill function in \cite{thattai,simpson,garder,elo} or an {\it
ad-hoc} function to fit the model to the experimental data in
\cite{blake,pnas08}). The approximation assumes that changes in the
levels of TF are reflected instantaneously in the transcription
rate. Although this approximation could be reasonable to study the
static deterministic behavior of transcriptional regulation
\cite{garder,elo}, it could leads to a significant underestimation
of transcriptional noise \cite{cox2}. Consequently, these models
cannot accurately describe how the overall regulatory process
affects noise expression. In this article, we propose a theoretical
model of transcriptional regulation that considers a CRS with
several regulatory binding sites for activating proteins. All
transition rates between CRS states follow the law of mass action
for elementary reactions. In this way, our model accounts for the
fact that the expression response is determined by the dynamics of CRS.

\section{The Model} We are interested in exploring how the molecular
interaction affect the cooperativity and the fluctuations level of
the gene expression. In this sense, we found that stronger
interaction between activators increases the level of noise
expression. In our model the transcriptional regulation is assumed
to be a stochastic process in which the regulatory system makes
transitions between different states. The model includes $N$
regulatory binding sites for the same TF (Fig. 1 illustrates the
case with three regulatory binding sites). The states
$s=1,2,\ldots,N+1$ represent, respectively, states with
$0,1,\ldots,N$ binding sites occupied by TFs. The states $s\ge N+2$
correspond to the transcriptional complex formation, where all
components required for transcription are assembled on the CRS. Once
the core transcription apparatus is formed, the synthesis of one
mRNA copy begins.

\begin{figure}[ht]
\centerline{\includegraphics[width=.75\textwidth]{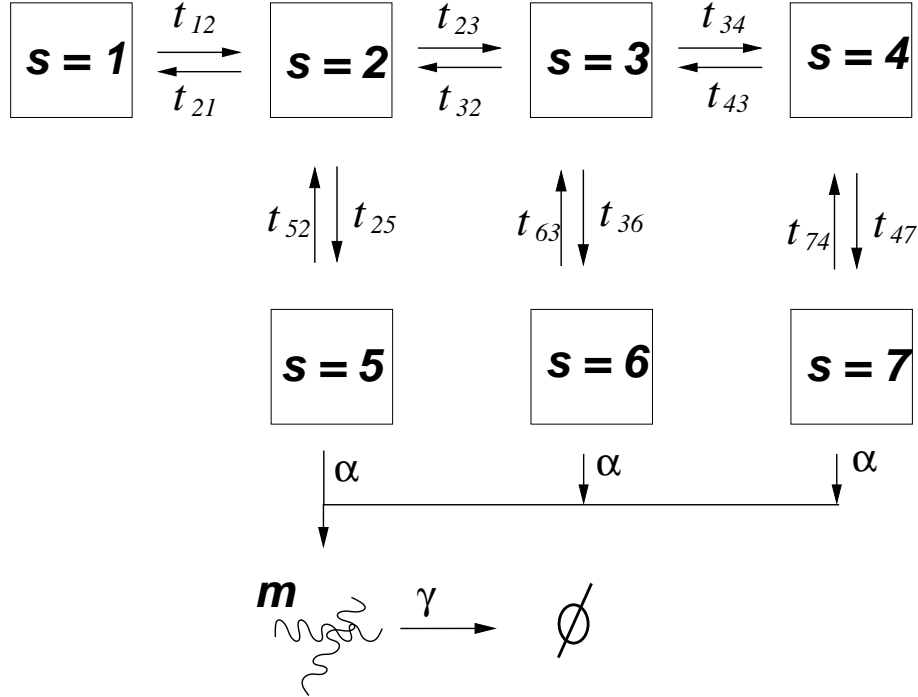}}
\caption{Kinetic regulatory scheme. All parameters shown in this
schematic diagram are constants.}\label{figu1}
\end{figure}

The working hypothesis is that TFs bound to DNA alter the
probability of transcriptional complex formation. Consequently,
states $s\le N+1$ are characterized by different rates of
transcriptional complex formation. For simplicity, we consider that
some bindings are sequential, i.e., TF does not bind or unbind after
transcriptional complex formation and transcriptional complex
does not assemble before TFs bound to DNA site. Additionally, we
consider that the sites are functionally identic, i.e., the model
does not distinguish among states with the same number of TFs bound
to regulatory binding sites. Thus, in our model, the states of CRS
are related more to the occupancy number rather than to the binding
status of each site. This additional simplification reduces the
number of states accessible to the CRS and allows us to explore the
role of cooperativity on noise expression without considering a
combinatorial number of states. The model assumes that mRNA is
synthesized at a rate which depend on the state $s$. mRNA is
considered to degrade linearly with rates $\gamma $.

As other authors \cite{thattai,elston}, we used the master equation
approach to derive the average behavior of mRNA level, as well as
its fluctuations. For this system, the state is specified by two
stochastic variables: the state of the CRS $s$, and the number of
transcripts $m$. We can write the probability to find, at any given
time $t$, the system in the state $(s,m)$ as a vector ${\bf
P}_{m}\left(t \right)= \left(P_{m}^1\left(t \right),P_{m}^2\left(t
\right),\ldots,P_{m}^{2N+1}\left(t \right) \right)$. The time
evolution for this probability is governed by the following master
equation:
\begin{eqnarray}
\label{master1} \nonumber \dot{P}_{m}^s &=&  \sum^{2N+1}_{r=1}
t_{s,r} P_{m}^r + \alpha _{s}\left(P_{m-1}^s - P_{m}^s \right)
\\ && + \gamma \left[ (m+1)P_{m+1}^s - m P_{m}^s
\right]
\end{eqnarray}
where $t_{s,r}$ are the elements of the transition matrix $\hat{T}$
and represent the transition probability per time unit from state
$r$ to state $s$, $\alpha _{s}$ are the elements of a vector and
correspond to the transcription rate of state $s$. The first term on
the right hand side of the equation (\ref{master1}) describes the
CRS dynamics, while the second and third terms correspond to the
production and degradation of mRNA,
respectively.

We are mainly interested on both the gene expression level and its
fluctuations. The first is measured through the first moment of the
number of transcripts $m$, and the last through the corresponding
variance, related to the second moment:
\begin{equation}\label{eq1}
\langle m\rangle = \sum_{m,r} m P_{m}^r, \ \ \ \ \ \sigma^2_m =
\langle m^2\rangle - \langle m\rangle^2,
\end{equation}
where the summation limits were suppressed for the sake of
readability. We want to remark that from now on, every sum over
transcript numbers will be from $m=0$ to $m=\infty$, while the sum
over CRS states will be from $r=1$ to $r=2N+1$.

Moments of $j$-th order can be written in term of its associated
partial moments
\begin{equation}%\label{eq1}
\langle m^j\rangle =\sum_r \langle m^j\rangle_r \ \ \ \ {\rm where}
\ \ \ \ \langle m^j\rangle_s = \sum_{m} m^j P_{m}^s,
\end{equation}
Note that the zero partial moments are the marginal probabilities
for the CRS to be in state $s$, regardless the number of
transcripts, i.e., $\langle m^0\rangle_s =P^s= \sum_{m} P_{m}^s $.

From Eqs. (\ref{master1}) we can derive a set of ordinary
differential equations for the time evolution of the partial moments
for any $j$. As there is no feedback, the system is linear and can
be solved analitycally. Thus, the time evolution of partial moments
for $j=0,1,$ and 2 are given by
\begin{eqnarray}
\label{j0} j=0 &&  \begin{array}{ll} \dot P^s = \sum_{r} t_{s,r}
P^{r}
\end{array}\\
\label{j1} j=1 &&
\begin{array}{lll}
\dot{ \langle m\rangle _s} &=& \sum_{r} t_{s,r} \langle m
\rangle_{r} + \alpha_{s} P^s - \gamma \langle m \rangle_s
\end{array} \\
\label{j2} j=2 &&
\begin{array}{lll}
\dot{ \langle m^2\rangle_s} &=& \sum_{r} t_{s,r} \langle
m^2\rangle_{r} - 2 \gamma \langle m^2\rangle _s + 2 \alpha _{s}
\langle m\rangle_s  + \gamma \langle m\rangle_s + \alpha _{s} P^s,
\end{array}
\end{eqnarray}
From these partial moments, we can readily find first order
differential equations governing the time evolution of the mean and
variance,
\begin{eqnarray}
\label{mediadot} \dot{\langle  m \rangle} &=& -\gamma \, \langle m
\rangle + \sum_r
\alpha _{r}\,  P^r \\
\label{vardot} \dot {\sigma_m^2} &=& -2\gamma  \, \langle m^2\rangle
+ \gamma
\, \langle m \rangle+ 2 \gamma \, \langle m\rangle ^2 \nonumber \\
&&+ \sum_r \alpha _{r} \, \left[ 2\langle m\rangle _r  + (1 - 2
\langle m\rangle) P^r  \right].
\end{eqnarray}
From Eq. (\ref{mediadot}), the steady-state solutions for the mean
value of $m$ is
\begin{equation}\label{eq3}
\langle m_*\rangle = \frac{1}{\gamma} \sum_{r} \alpha _{r} P^r_*,
\end{equation}
where $_*$ denotes the steady state solution. The steady-state
solution of the probability vector ${\bf P_*}$ corresponds to the
normalized eigenvector associated to the zero eigenvalue of the CRS
transition matrix $ \hat T {\bf P_*} = {\bf 0}$.

The steady state solution for the variance follows from Eq.
(\ref{vardot}) \begin{equation}\label{eq4} \sigma^{2*}_m = \langle
m_*\rangle - \langle m_*\rangle^2 + \frac{1}{\gamma} \sum_{r} \alpha
_{r} \langle m_*\rangle_r, \end{equation} where $\langle m
_*\rangle_r $ is determined as the solution of the linear equation
\begin{equation}\label{eq5} \sum_r \left( t_{s,r}- \gamma
\delta_{s,r}\right)\langle m_*\rangle_r=-\alpha_{s} P^s_*,
\end{equation} where $\delta_{s,r}$ is the Kronecker delta. The
expressions (\ref{eq3}--\ref{eq5}) are general, in the sense that
they are valid for $N$ binding sites in the CRS. In this paper, we
have limited the study to the case of Fig. 1, i.e., a CRS with
$N=3$, and $\alpha _{s}=\alpha $ for $s\ge 4$ and zero otherwise.
We are motivated to set $N=3$ because the cooperative effects are
more apparent for greater $N$. However, an approximation used in
the next Section, could not be adequate for higher $N$.
The TFs can bind to regulatory sites with a probability proportional
to TF concentration $c$ ($ck_{12},ck_{23}$ and $ck_{34}$, where
$k_{i,j}$ are the kinetic rates), following the law of mass action
for elementary reactions. TF unbinding events depend only on the
kinetic constants ($k_{21},k_{32}$ and $k_{43}$). In this case the
transition matrix $\hat{ T} $ can be written as
\begin{equation}\label{matrix} \hat{ T}= \left(
  \begin{array}{cccccccc}
    -c k_{12} & k_{21} & 0 & 0 & 0 & 0 & 0 \\
    ck_{12} & -(k_{21}+c k_{23}+k_{25}) & k_{32} & 0 &  k_{52} & 0 & 0 \\
    0 & c k_{23} & -(k_{32}+c k_{34}+k_{36}) & k_{43} & 0 &  k_{63} & 0 \\
    0 & 0 & c k_{34} & -(k_{43}+ k_{47}) & 0 & 0 &  k_{74} \\
    0 & k_{25} & 0 & 0 &  -k_{52} & 0 & 0 \\
    0 & 0 & k_{36} & 0 &  0 & -k_{63} & 0 \\
    0 & 0 & 0 & k_{47} &  0 & 0 & -k_{74} \\
  \end{array} \right)
\end{equation}
and the associated steady state solutions of the partial
probabilities $P^s_*$ involved on Eq. (\ref{eq3}) are
\begin{eqnarray}
\nonumber P_*^5= \frac{c
K_2K_5}{1+c\left(K_2\left(K_5+1\right)\right)+c^2\left(K_2K_3\left(K_6+1\right)\right)
+c^3\left(K_2K_3K_4\left(K_7+1\right)\right)}\\
\nonumber P_*^6= \frac{c^2
K_2K_3K_6}{1+c\left(K_2\left(K_5+1\right)\right)+c^2\left(K_2K_3\left(K_6+1\right)\right)
+c^3\left(K_2K_3K_4\left(K_7+1\right)\right)}\\
P_*^7= \frac{c^3
K_2K_3K_4K_7}{1+c\left(K_2\left(K_5+1\right)\right)+c^2\left(K_2K_3\left(K_6+1\right)\right)
+c^3\left(K_2K_3K_4\left(K_7+1\right)\right)}
\end{eqnarray}
where $K_s=\frac{k_{s-1,s}}{k_{s,s-1}}$ for $s=2,3,4$ and
$K_s=\frac{k_{s-3,s}}{k_{s,s-3}}$ for $s=5,6,7$. Replacing these
expressions on Eq. (\ref{eq3}), we find the explicit form for the
steady state expression level of transcripts
\begin{eqnarray}
\label{hill} m^* = \frac{\alpha}{\gamma} \frac{K_2 K_5 c + K_2 K_3
K_6 c^2 + K_2 K_3 K_4 K_7 c^3}{1+  K_2 (1+ K_5) c +  K_2 K_3 (1+K_6)
c^2 +  K_2 K_3 K_4 (1+K_7) c^3}
\end{eqnarray}
Unfortunately, though a closed expression for the variance is
obtained it is too long to report here. From this point the $_*$
will be suppressed from the steady states expressions.

\section{Binding cooperative mechanisms}
As the regulatory sites are assumed to be functionally identic, we
can introduce a relationship between the TF binding/unbinding when
there is no interaction between the TFs. Thus, if the probability
per time unit that a single TF molecule binds to a regulatory site
is $p$, we have $ k^{\rm o}_{12}=3p$, $k^{\rm o}_{23}=2p$, $k^{\rm
o}_{34}=p$, where $^{\rm o}$ indicates that there is no interaction
between TFs. Similarly, unbinding rates are given by $ k^{\rm
o}_{21}=q$, $k^{\rm o}_{32}=2q$, $k^{\rm o}_{43}=3q$, where $q$ is
the probability per time unit that a single TF molecule unbinds from
an occupied site.

Now, we will assume that the probability for a TF molecule to bind
to a given regulatory site arises from: i) the free energy of
binding a TF to the specific site $\Delta G_{\rm DNA}$, ii) the free
energy of interaction between TF molecules bound to adjacent sites
$\Delta G_{\rm I}$. From these assumptions and the principle of
detailed balance \cite{hill85}, we are able to
find a relationship between kinetic rates with and without
interactions among TFs. In the case $\Delta G_{\rm I}=0$, we have
\begin{equation}ck^{\rm o}_{s,s+1}/k^{\rm
o}_{s+1,s}=e^{-\frac{\Delta G_{\rm DNA}}{RT}}  \ \ \ \ \ {\rm for} \
\ s=1,2  \ {\rm and} \ 3, \end{equation} where $k^{\rm o} _{s,s+1}$
represents the transition rate from state $s$ to state $s+1$ when
there is no interaction between TFs ($k^{\rm o}_{s+1,s}$ represents
the rate for reverse transition). In general, the TF molecules
interact with each other, i.e., $\Delta G_{\rm I}<0$. Consequently,
we have \begin{equation}\label{rel3}
\frac{k_{s,s+1}}{k_{s+1,s}}=\varepsilon^{a_s} \frac{k^{\rm
o}_{s,s+1}}{k^{\rm o}_{s+1,s}}, \end{equation} where
$\varepsilon =e^{-\frac{\Delta G_{\rm I}}{RT}} $ represents the intensity of the
interaction between TFs. $a_s$ represents the number of interactions.
We assume that all bounded TFs interact with the new one, regardless
of their position on the CRS, thus we have $a_s=s-1$. This means for
example that in the state $s=4$, the third TF interacts with the two
bounded TFs. As the cooperative effects are more apparent for
greater $N$, we are motivated to use a high number of binding
sites. However, the assumption $a_s=s-1$ could be inadequate for
$s>4$ when $N>3$ due to the greater separation between faraway
sites. For this reason, we set $N=3$ for further calculations.
The relationship (\ref{rel3}) leaves an extra degree of freedom,
because the interaction between TFs can increases the binding rate
$k_{s,s+1}$, increasing the ability for new TF recruitment for DNA
binding, or it can diminishes the unbinding rate $k_{s+1,s}$,
increasing the stability of the TF-DNA bound. The first case will be
denoted here as the recruitment mechanism (RM), while the second
case will be denoted as stabilization mechanism (SM). These two
mechanisms are not mutually excluding, and they could be acting
simultaneously in real life, but in order to study their effect on
the regulatory response and its associated fluctuations, we will
consider the alternative cooperativity binding mechanisms
separately. Thus, using relations (\ref{rel3}) and the relations for
binding/unbinding rates, we obtain \begin{eqnarray}\label{rel4}
{k_{s,s+1}}&=&\varepsilon^{\left (s-1 \right)} \left(4-s\right) p
\nonumber \\ {k_{s+1,s}}&=& s q, \end{eqnarray} for the first
mechanism, while for the second mechanism we have
\begin{eqnarray}\label{rel5} {k_{s,s+1}}&=& \left(4-s\right) p,
\nonumber \\ {k_{s+1,s}}&=&\varepsilon^{\left (1-s \right)} s q.
\end{eqnarray} Table I summarizes the parameter values used in this
work (the time unit is min and concentration is an arbitrary unit.)
Binding and unbinding parameters were obtained considering $p=0.25$,
$q=0.75$ and $\varepsilon=6$ (which is equivalent to $\Delta G_{\rm
I}\simeq -1.0$ Kcal). Cases RM and SM have the same TF interaction
intensity, while $\varepsilon=0$ when there is no interaction
between TFs.

\begin{table}[ht]
\begin{tabular}{l|c|c|c|l|c} \hline
\multicolumn{4}{c|}{\underline{TF binding/unbinding rates} }
& \multicolumn{2}{|c}{\underline{TC formation rates}} \\
& RM case & SM case & $\varepsilon _0$ case & \multicolumn{2}{|c}{}   \\
\hline
$k_{12}$ & 0.75& 0.75 & 0.75&  $k_{25}$ &0.50 \\
$k_{21}$ & 0.75& 0.75 & 0.75&  $k_{52}$ &0.50 \\
$k_{23}$ & 3.00& 0.50 & 0.50&  $k_{36}$ &1.00 \\
$k_{32}$ & 1.50& 0.25 & 1.50&  $k_{63}$ &0.50 \\
$k_{34}$ & 9.00& 0.25 & 0.25&  $k_{47}$ &1.50 \\
$k_{43}$ & 2.25& 0.0625& 2.25&  $k_{74}$ &0.50\\
\hline \hline
\multicolumn{6}{c}{ \ \ \ \ \ \ \ \underline{Production and degradation rates}} \\
\multicolumn{2}{c|}{mRNA (layer III)} &$\alpha $ & 1.50 & $\gamma $& 0.03 \\
\hline
\end{tabular}
\caption{Kinetic parameters values. RM case: recruitment mechanism;
SM case: stabilization mechanism; $\varepsilon _0$ case: there is no
interaction between TFs. The time unit is min and the concentration
is an arbitrary unit.}
\end{table}

\section{Results} From the Eq. (\ref{hill}) for the mean, Eqs.
(\ref{eq4}-\ref{eq5}) for the variance, and using the parameters
values of Table I, we study the response of an activator switch. We
consider that the kinetic rates for transcriptional complex
formation increase linearly with the occupancy number, i.e.
$k_{s,s+3}\propto s $. With this condition we assume that there is
no synergism between TF and transcriptional complex formation
\cite{syner}. This synergism can contribute to the effective
cooperativity (data not shown). Figure 2A depicts the average number
of mRNA copies $\langle m\rangle$ and Figure 2B depicts the standard
deviation $\sigma _m$ as a function of the activator concentration $c$,
obtained analytically for the three cases: RM case, the
interaction between TFs increases the binding rates, (solid line);
SM case, the interaction between TFs decreases the unbinding rates;
and ($\varepsilon_0$) case, where there is no interaction between
TFs (dotted line). The mean response of RM and SM cases are exactly
the same. \begin{figure}[ht]
\centerline{\includegraphics[width=.75\textwidth]{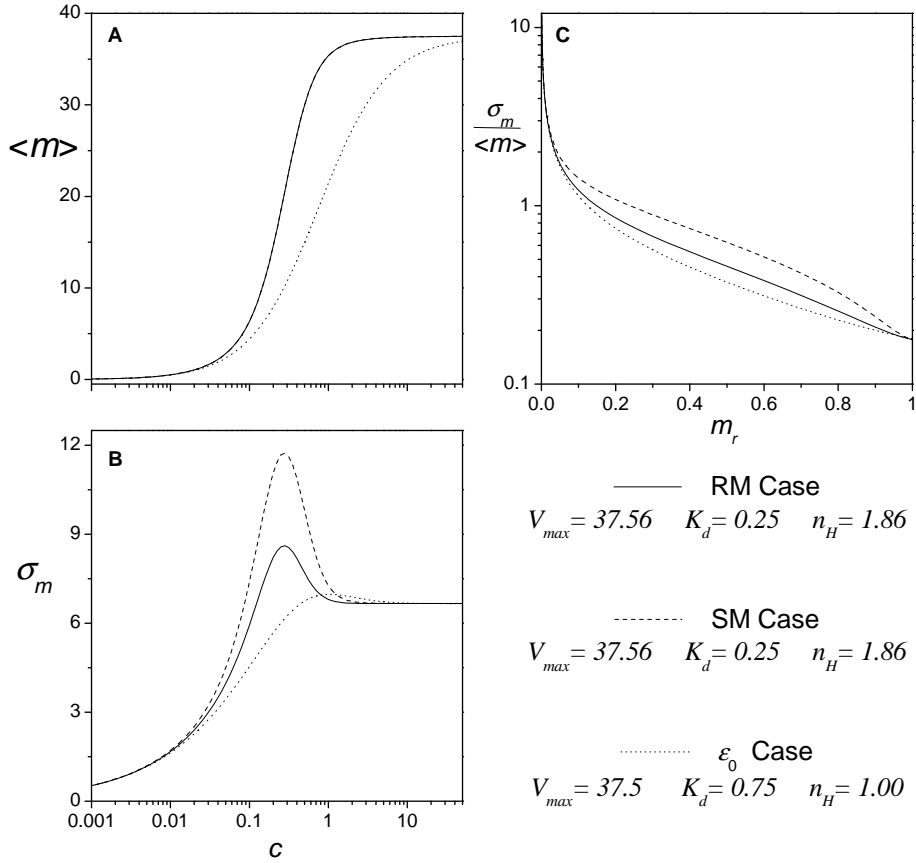}}
\caption{Average number of mRNA copies as a function of $c$ in
steady state for three different cases: RM (solid line), SM
(overlapping the former case dashed line), and $\varepsilon _0$ case
(dotted line). See Table I for parameter values. (B) Associated
standard deviations of the transcript number as a function of $c$
(C) Associated noise versus the transcriptional
efficiency.}\label{figu2} \end{figure} The regulatory function for
both examples with cooperative binding fit the Hill function
$f\left(c\right)=V_{max}/\left[ 1+ \left( c/K_d\right)^{n_h}
\right]$, where $c$ is the binding protein concentration, $V_{max}$
is the maximum number of transcripts, $K_d$ is the dissociation
constant, and $n_h$ is known as the Hill coefficient which
determines the steepness of the regulatory function. Both RM and SM
cases the Hill coefficient is $1.86$, while the $\varepsilon_0$ case
is associated to a response without cooperativity $n_h=1.00$. In all
cases the fluctuation level estimated by the standard deviation
$\sigma _m$ (see Figure 2B) has a peak centered around the
dissociation constant $K_d$. Relative fluctuations are characterized
by the normalized standard deviation. Analyzing an un-normalized
measure can lead to artefactual results \cite{paul04}. Fig. 2C
depicts the noise $\eta$ (defined as $\sigma _{m}/\langle m\rangle$)
as a function of the transcriptional efficiency $m_r$ (defined as
the ratio between transcription and maximum transcription $\langle
m\rangle /\langle m\rangle _{max}$, where $\langle m\rangle _{max}$
is the maximum of $\langle m\rangle$) for each case in Figure 2A.

Figure 2 shows that cooperativity has an effect not only in
controlling the expression response increasing $n_h$, but also in
increasing the relative size of fluctuations. We note also that
though the regulatory function in RM and SM cases is the same, the
mechanism of increasing the TF-DNA complex stabilization (SM) is
associated to higher level of noise (dashed line) than the mechanism
involving an improvement in the recruitment ability of new TF to the
DNA (RM) (solid line). Figures 2 suggest that the two cooperative
binding mechanisms considered here affect the fluctuation level in a
differential way but not the regulatory function. Even though this
function is altered by the FT interactions, it is not possible to
distinguish between the alternative mechanisms through the
regulatory function only.

\begin{figure}%[ht]
\begin{tabular}{cc}
\includegraphics[width=.5\textwidth]{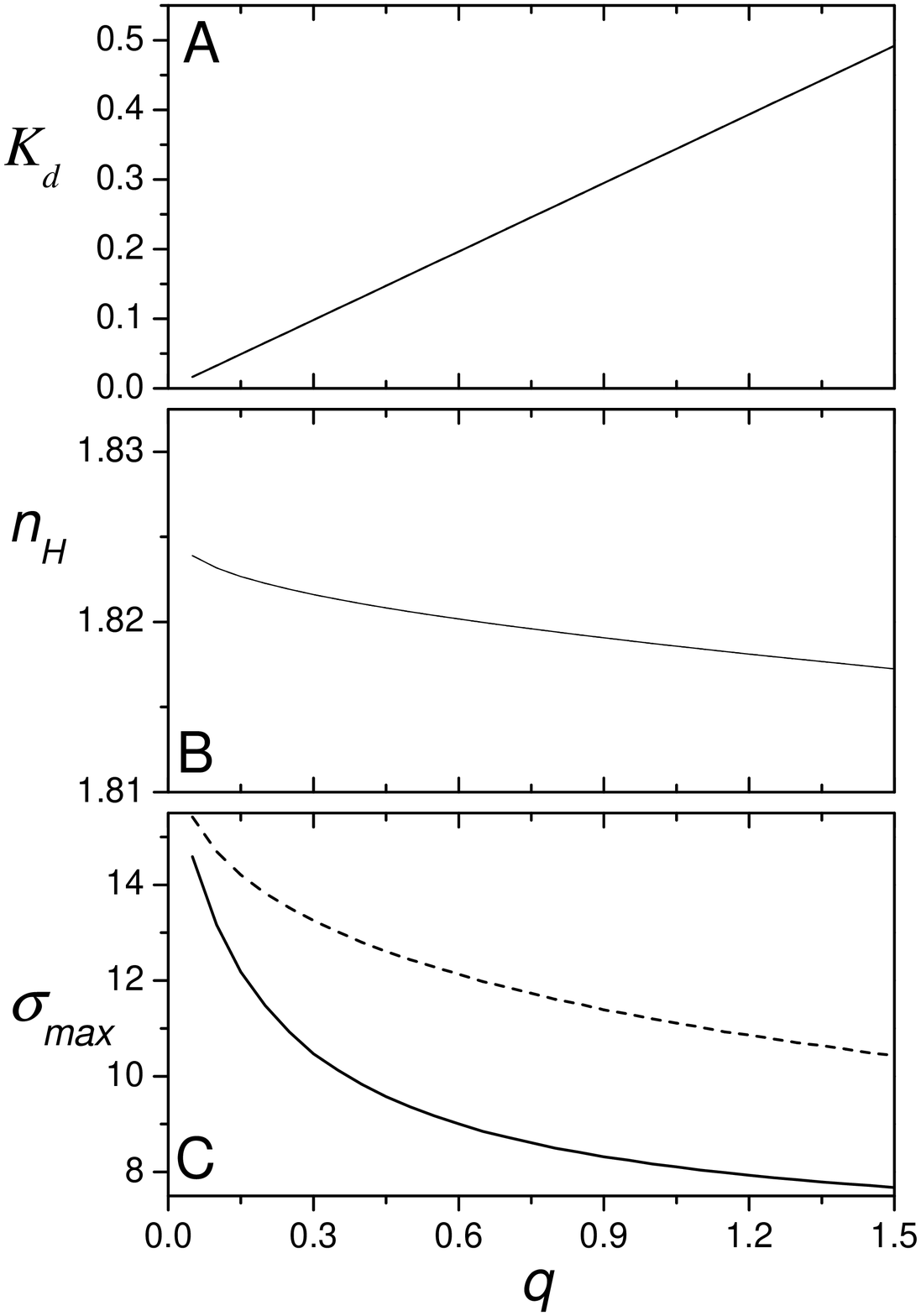}&
\includegraphics[width=.5\textwidth]{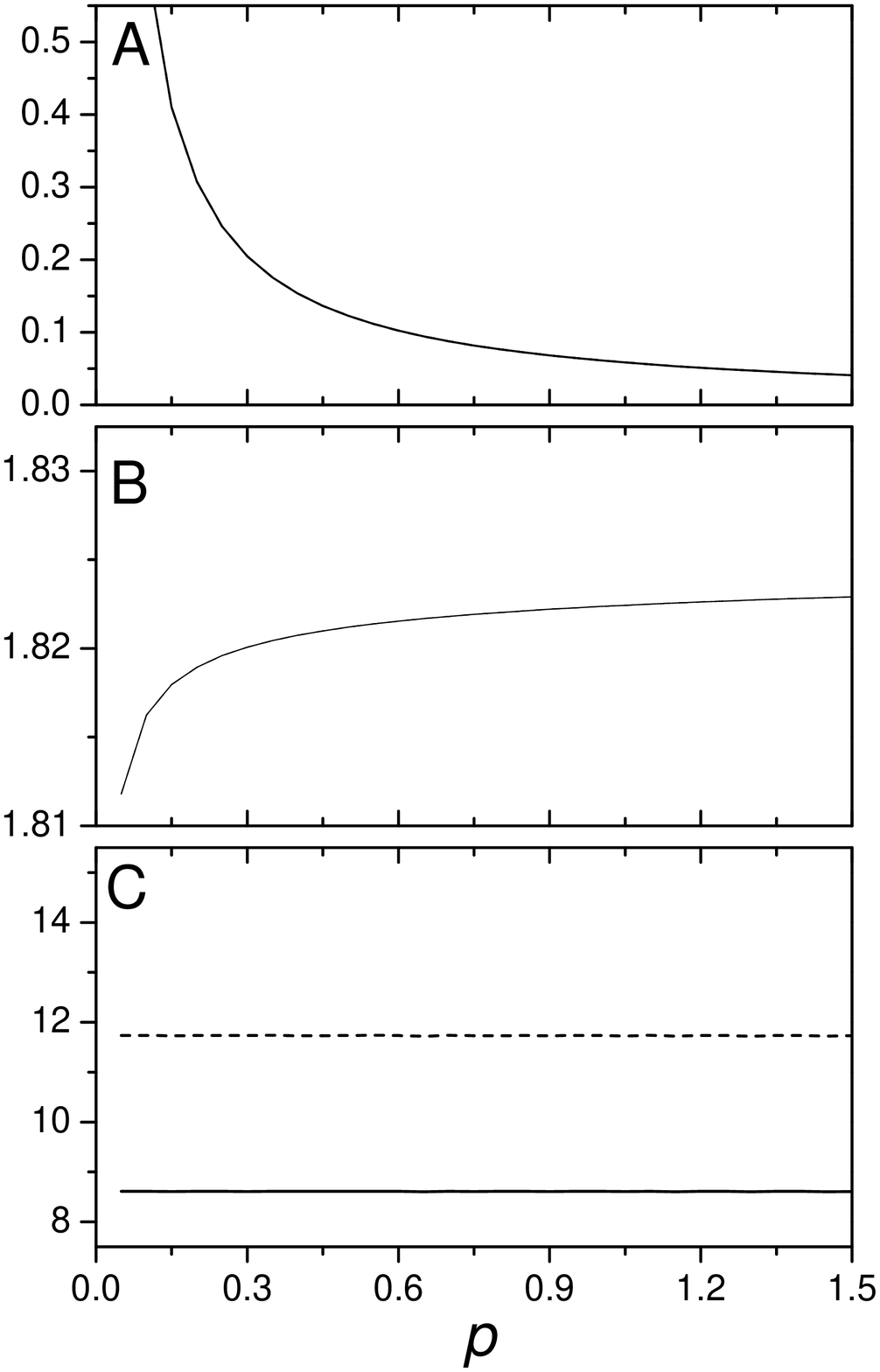}
\end{tabular}
\caption{Half maximal value of the mean response $K_d$ as a function
of the unbinding rate $q$, and as function of the binding rate $p$.
(B) Hill coefficient $n_h$ as a function of the unbinding rate $q$
and as function of the binding rate $p$. (C) Maximum value that
reaches the standard deviation {\it vs} the unbinding rate $q$, and
as function of the binding rate $p$. Left: $q$ varies keeping
$p=0.25$, Right: $p$ varies keeping $q=0.75$. For RM mechanism
(solid line) and SM mechanism (dashed line).} \label{figu3}
\end{figure}

We also analyze how the binding and unbinding rates, $p$ and $q$
respectively, affect the regulatory function and the fluctuation
level. Figure 3A depicts the behavior of the dissociation constant
$K_d$, as function of the unbinding rate $q$ (left) and as function
of binding rates $p$ (right), keeping the other rates constant.
$K_d$ does not depend on which cooperative binding mechanism is
acting (the curves are completely overlapped). However, $K_d$
increases linearly with binding rate $q$, and it is inversely
proportional to the binding rate $p$. Figure 3B depicts the behavior
of the Hill coefficient $n_h$, as function of the binding/unbinding
rate (right/left), keeping the other rates constant. We can observe
that $n_h$ is almost independent on these rates. From Fig. 3C, we
note that the level of noise is sensitive to the type of cooperative
binding mechanisms which is. $\sigma _{max}$ decreases with the
unbinding rate more slowly in the SM (dashed line) than in the RM
(solid line). The difference between the two mechanisms diminishes
when the unbinding rate decreases, while the maximum value of
dispersion is not affected when the unbinding rate $p$ varies. Due
to the complexity of the problem are not able to provide a
quantitative proof that SM leads to larger noise than the RM in all
conditions. However, from figure 3C we can see that noise level does
not depend on the binding rate $p$, but it depends on the the
unbinding rate $q$. As the interaction energy between TFs decreases
the unbinding rate in the SM, it is expected that SM has associated
a higher level of fluctuation than the RM.

Finally, we show how the interactions between TFs alters both $K_d$
and $n_h$ parameters of the regulatory function and also the
fluctuation level. Figure 4A illustrates how the parameters $K_d$
and $n_h$ are affected by the interaction intensity $\varepsilon$.
The Hill coefficient (filled circle symbols), scaled on the right
vertical axis, increases with $\varepsilon$, suggesting that the
steepness of the regulatory function depends linearly on the free
energy $\Delta G_{\rm I}$. Furthermore, the dissociation constant
$K_d$ (open square symbols), scaled on the left vertical axis of
Figure 4A, decreases with the interaction intensity. Finally, we
found that the fluctuation level increases with the interaction
intensity. Figure 4B depicts the maximum value of the standard
deviation $\sigma _{max}$ as a function of $\varepsilon$. We have
observe that the RM (solid line) is less sensitive to the
interaction intensity than the SM (dashed line). We want to remark
that the differences between the recruitment and stabilization
mechanisms vanish when there is not interaction energy between TF
($\Delta G_I=0$). These observation suggest that our model predicts
that interactions between TFs improve the response of the regulatory
system in the sense of specificity (higher $n_h$) and sensitivity
(lower $K_d$). But, in contrast, the system loses accuracy because
the noise increases with the intensity of the interaction.

\begin{figure}%[ht]
\centerline{\includegraphics[width=.75\textwidth]{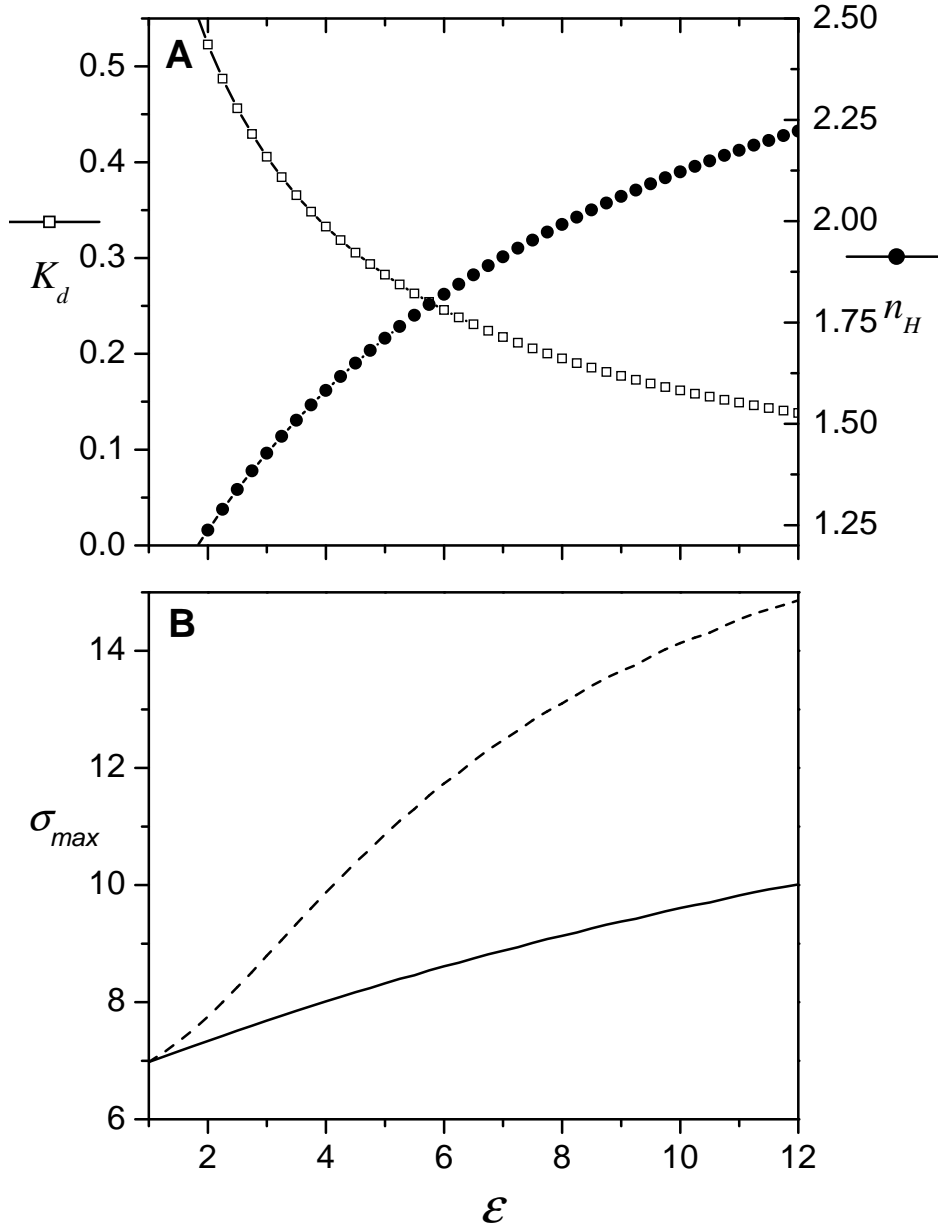}}
\caption{(A) Half maximal value $K_d$ of the mean response as a
function of $\varepsilon$ (scale on left axis). Hill coefficient
$n_h$ as a function of $\varepsilon$ (scale on right axis). The
curves associated to RM and SM are completely overlapped. (B)
Maximum value that reaches the standard deviation $\sigma _{max}$ as
a function of $\varepsilon$. Solid line corresponds to the
recruitment mechanism, while dashed line corresponds to the
stabilization mechanism.} \label{figu4}
\end{figure}

\section{Conclusion}
We have shown that a model which includes several binding sites is
able to address the question of cooperative binding effects on
fluctuations. The first moment of $m$ is the same as that obtained
from thermodynamic models, which depends solely on equilibrium
constants. Nevertheless, second moments have allowed us to introduce
new quantitative insights on the TF cooperative binding effects in
the cell-to-cell variability. We found that two different cooperative
binding mechanisms can be distinguished: the RM which increases the
ability for new TF recruitment, and the SM which increases the stability
of the TF-DNA bound. In both mechanisms, the Hill coefficient and level
of noise increase as the interaction energy between activators increases.
Only a few kilocalories of binding energy between TFs have a dramatic
effect on the noise level, which also depends on the acting cooperative
binding mechanisms. The other hand, the mechanism that reduces the
unbinding rates is associated to a greater level of noise which is
in agreement with two state model \cite{kaern}.

Both mechanisms reported here are derived from the thermodynamics
relationship used in the modeling.
This cannot be done in simpler models that use regulatory expression
function rather than TFs thats bound to several binding sites on DNA
following the law of mass action. These different mechanisms have
not been reported previously. Although the proposed model is more
complicated than previous, it can also be solved analytically. Thus,
the model constitutes an adequate frame to discuss the impact of the
diverse cooperativity mechanisms on the gene expression
fluctuations. However, we want to remark that the presented model is
limited to intrinsic contribution of noise, i.e. it does not regard
the fluctuation on the TF concentration and other extrinsic source
of noise, which certainly contribute to the total noise.

\section*{Acknowledgements}
We thank Christina McCarthy for critical review of the manuscript.
P.S.G. thanks CONICET for financial support. L.D. is researcher of
CONICET (Argentina). D.M. is researcher of CICPBA (Buenos
Aires-Argentina).

\end{document}